\documentclass[11pt,leqno]{article}

\usepackage{graphicx} 

\usepackage[dvipsnames]{color} 
\definecolor{Red}{named}{Red}

\usepackage{amsmath} 
\usepackage{latexsym,epsfig,bm,amssymb} 
\usepackage{color} 
\usepackage{amsthm,mathrsfs}

\newcommand{\nc}{\newcommand} 
\nc{\dps}{\displaystyle} 

\newtheorem{theorem}{Theorem}

\newcommand{\loota}{\hbox{\enspace{\vrule height 7pt depth 0pt width 
      7pt}}} 
\nc{\RR}{\mbox{\rm I$\!$R}} 
 
 \newcommand{\beqn}{\begin{eqnarray}} 
 \newcommand{\eeqn}{\end{eqnarray}} 
 \newcommand{\be}{\begin{equation}} 
 \newcommand{\ee}{\end{equation}} 
 \newcommand{\ba}{\begin{array}} 
 \newcommand{\ea}{\end{array}} 
 \newcommand{\pa}{\partial} 
 \newcommand{\re}{\ref} 
 \newcommand{\ci}{\cite} 
 \newcommand{\ds}{\displaystyle} 
 \newcommand{\la}{\label} 
 \newcommand{\bfr}{\begin{flushright}} 
 \newcommand{\efr}{\end{flushright}} 
  
 \newcommand{\rIm}{{\rm Im\5}} 
 \newcommand{\rRe}{{\rm Re\5}} 
\newcommand{\bfl}{\begin{flushleft}} 
\newcommand{\efl}{\end{flushleft}} 
\newcommand{\fr}{\frac}

\newcommand{\ov}{\overline} 
\newcommand{\st}{\stackrel}

\newcommand{\toLp}{\st{L^p}\longrightarrow} 
\newcommand{\toHd}{\st{H^2}\longrightarrow} 
\newcommand{\toLd}{\st{L^2}\longrightarrow} 

\newcommand{\toLpR}{\st{L^p_R}{\longrightarrow}} 
\newcommand{\toLLR}{\st{L^2_R}{\longrightarrow}}

\def\longrightharpoonup{\relbar\joinrel\rightharpoonup} 
\newcommand{\tow}{\st{L^2_w}{\longrightharpoonup}}

\newcommand{\rot}{{\rm rot\5\5}}

\newcommand{\bo}{{\hfill\loota}} 
\newcommand{\supp}{{\rm supp\5}} 
\renewcommand{\Pr}{{\bf Proof.~}}

\newcommand{\bba}{{\bf a}} 
 
\newcommand{\n}{{\bf n}} 
\newcommand{\bb}{{\bf b}} 
\newcommand{\e}{{\bf e}}

\newcommand{\x}{{\bf x}} 
\newcommand{\y}{{\bf y}}

\newcommand{\bk}{{\bf k}} 
 
\newcommand{\cm}{{\rm m}}

\newcommand{\E}{{\cal E}}

\newcommand{\cO}{{\cal O}}

\newcommand{\ccT}{{\cal T}} 
\newcommand{\cS}{{\cal S}}

\newcommand{\ve}{\varepsilon} 
\newcommand{\vp}{\varphi} 
 
\newcommand{\De}{\Delta} 
\newcommand{\de}{\delta} 
 
\newcommand{\al}{\alpha} 
 
\newcommand{\Ga}{\Gamma} 
\newcommand{\si}{\sigma} 
\newcommand{\om}{\omega} 
 
\newcommand{\na}{\nabla} 
\newcommand{\Si}{\Sigma} 
\newcommand{\lam}{\lambda} 
\newcommand{\Lam}{\Lambda} 
 \newcommand{\h}{{\hbar}}

\newcommand{\5}{{\hspace{0.5mm}}} 
\newcommand{\C}{{\mathbb C}} 
\newcommand\R{{\mathbb R}} 
\newcommand\Z{{\mathbb Z}} 
\newcommand\T{{\mathbb T}}

 \renewcommand{\theequation}{\thesection.\arabic{equation}} 
\renewcommand{\thesection}{\arabic{section}} 
\renewcommand{\thesubsection}{\arabic{section}.\arabic{subsection}} 
\newtheorem{qtheorem}{QTheorem}[section] 
 
\renewcommand{\thetheorem}{\arabic{section}.\arabic{theorem}} 
\newtheorem{defin}[theorem]{Definition} 
 
\newtheorem{lemma}[theorem]{Lemma} 
\newtheorem{example}[theorem]{Example} 
\newtheorem{exercice}[theorem]{Exercise} 
\newtheorem{remark}[theorem]{Remark} 
\newtheorem{remarks}[theorem]{Remarks} 
\newtheorem{cor}[theorem]{Corollary} 
\newtheorem{pro}[theorem]{Proposition} 
 
\newtheorem{coms}[theorem]{Comments}

\newcommand{\bd}{\begin{defin}} 
 \newcommand{\ed}{\end{defin}} 
\newcommand{\bt}{\begin{theorem}} 
 \newcommand{\et}{\end{theorem}} 
\newcommand{\bqt}{\begin{qtheorem}} 
 \newcommand{\eqt}{\end{qtheorem}}

\newcommand{\bp}{\begin{pro}} 
 \newcommand{\ep}{\end{pro}} 
 
\newcommand{\bl}{\begin{lemma}} 
 \newcommand{\el}{\end{lemma}} 
\newcommand{\bc}{\begin{cor}} 
 \newcommand{\ec}{\end{cor}} 
 
\newcommand{\bex}{\begin{example}} 
 \newcommand{\eex}{\end{example}} 
\newcommand{\bexs}{\begin{examples}} 
 \newcommand{\eexs}{\end{examples}}

\newcommand{\bexe}{\begin{exercice}} 
 \newcommand{\eexe}{\end{exercice}}

\newcommand{\br}{\begin{remark} } 
 \newcommand{\er}{\end{remark}} 
\newcommand{\brs}{\begin{remarks}} 
 \newcommand{\ers}{\end{remarks}}

\newcommand{\bcoms}{\begin{coms}} 
\newcommand{\ecoms}{\end{coms}}

\begin{document} 
 
% \hspace{50mm} submitted to 
%{\it SIAM Journal on Mathematical Analysis., 2013}
%\medskip\medskip\medskip

% \hspace{50mm} arXiv: 1310.3084v1 [math-ph] 11 Oct 2013

~ \vspace{20mm} 
\begin{center} 
%\hspace{10cm} 
%{\huge THIS IS DRAFT} 
%\\~ 
%\\ 
{\huge\bf  On  crystal ground state in 
\bigskip\\ 
the Schr\"odinger--Poisson model} 
\\~ 
\\~ 
\\ 
\vspace{15mm} 
{\large A.\,I.~Komech 
\footnote{Supported partly by 
Alexander von Humboldt Research Award, 
Austrian Science Fund: P22198-N13, 
and grants of DFG and the Russian Foundation for Basic Research.}}\\ 
{\it Faculty of Mathematics of  Vienna  University and\\ 
Institute for Information Transmission Problems RAS\\} 
\bigskip\bigskip\bigskip\bigskip\bigskip

\end{center} 
 
\begin{abstract} 
A~space-periodic ground state is shown to exist for lattices of
smeared
ions  in $\R^3$ coupled to the Schr\"odinger and scalar fields. 
The  elementary cell is necessarily neutral. 
 
The 1D, 2D and 3D lattices in $\R^3$ are considered, and 
a~ground state is constructed by minimizing the energy per cell. 
The case of a~3D lattice is rather standard,  because 
the elementary cell is compact, and the spectrum of the Laplacian is discrete. 

In the cases of  1D and 2D lattices, the energy functional 
is differentiable only on a dense set of variations, 
due to the presence of the continuous spectrum 
of the Laplacian that causes the infrared divergence of the Coulomb bond. 
Respectively,  
the construction of electrostatic potential 
and
the derivation of the Schr\"odinger equation for the minimizer 
 in these cases require an extra argument.

 The space-periodic ground states for 1D and 2D lattices 
 give the model of
 the nanostructures similar to 
 the carbon nanotubes and graphene respectively.

\end{abstract}

%%\tableofcontents 

%%%%%%%%%%%%%%%%%%%%%%%%%%%%%%%%%%%%%%%%%%%%%%%% 
%%%%%%%%%%%%%%%%%%%%%%%%%%%%%%%%%%%%%%%%%%%%%% 
 
 \newpage 
\setcounter{equation}{0} 
\setcounter{section}{-1} 
\setcounter{section}{0} 
\section 
{Introduction} 
We consider  $d$-dimensional ion lattices in $\R^3$, 
\be\la{Ga3} 
\Ga_d:=\{\x(\n)=\bba_1 n_1+\dots+\bba_d n_d: \n=(n_1,...,n_d)\in\Z^d  \}, 
\ee 
where $d=1,2,3$ and $\bba_k\in\R^3$ are linearly independent periods. 
A~2D lattice (respectively, 1D lattice) is a mathematical model of a~monomolecular film (a~wire). 
%% similar to nanostructures like the graphene (the~carbon nanotube).

Born and Oppenheimer \ci{BO} developed 
the  quantum dynamical approach to the crystal structure, 
separating the motion of  `light electrons' and of `heavy ions'. 
As an extreme form of this separation, 
the ions could be 
considered as classical nonrelativistic particles governed by the 
Coulomb force, 
while the electrons could be  described by the Schr\"odinger 
equation neglecting the electron spin. 
The scalar potential 
is the solution to the corresponding Poisson equation. 

We consider the crystal with $N$ ions per cell.
Let us denote  by $\mu_j$
the charge density of an ion and by $M_j>0$ its mass, $j=1,...,N$.
Then the coupled equations 
 read 
\beqn 
i\h\dot\psi(\x,t)&=&-\fr{\h^2}{2\cm}\De\psi(\x,t)+e\phi(\x,t)\psi(\x,t), ~~~~~\x\in\R^3,
\la{LPS1} 
\\ 
\nonumber\\ 
\qquad \Bigl [\ds\fr1{c^2}\pa_t^2-\De\Bigr] 
\phi(\x,t)&=&\rho(\x,t):= 
\sum_{j=1}^N\5\sum _{\n\in\Z^d} 
\mu_j(\x-\x(\n)-\x_j(\n,t))+e|\psi(\x,t)|^2, ~~~~~\x\in\R^3,\la{LPS2} 
\\ 
\nonumber\\ 
M_j\ddot\x_j(\n,t) 
&=&-(\na\phi(\x,t),\mu_j(\x-\x(\n)-\x_j(\n,t))),
\quad \n\in\Z^d,\quad 
j=1,\dots,N. 
\la{LPS} 
\eeqn 
Here 
$e<0$ is the electron charge, 
$\cm$ is its mass, 
$\psi(\x,t)$ denotes the 
wave function of the electron field, and 
$\phi(\x,t)$ is the electrostatic  potential 
generated by the ions and the electrons. 
Further, 
$(\cdot,\cdot )$ 
stands for the scalar product in the Hilbert 
space $L^2(\R^3)$. 
All derivatives here and below 
are understood in the sense of distributions. 
The system is  nonlinear and translation invariant, 
i.e., $\psi(\x-\bba,t)$, $\phi(\x-\bba,t)$, 
$\x_j(\n,t)+\bba$ is also a~solution for any $\bba\in\R^3$ . 
 
%%%%%%%%%%%%%%%%%
%%%%%%%%%%%%%%%%%%%%%
 A dynamical quantum 
 description of the solid state as many-body system
is not rigorously established yet
 (see Introduction  of  \ci{LAK} and Preface of \ci{Peierls}).
 %%%%%%%%%%%%%%%%%%%%%
Up to date rigorous results concern only the ground state in different models
(see below).

The classical "one-electron" theory of Bethe-Sommerfeld,  
based on  periodic Schr\"odinger equation, does not take into account
oscillations of  ions.
Moreover, the choice of the periodic potential in this theory is very problematic, and 
corresponds to a fixation of the ion positions which are unknown.
 
The system (\re{LPS1})--(\re{LPS})  eliminates these  difficulties
though  it does  not respect the electron spin like the periodic Schr\"odinger equation.
To remedy this deficiency we should replace the Schr\"odinger equation by 
the Hartree--Fock equations as the next step to more realistic model.
However, we expect that the  techniques developed for the system
(\re{LPS1})--(\re{LPS})
will be useful also for more realistic  dynamical models of crystals.
These goals were our main motivation in writing this paper.
%%%%%%%%%%%%%%

 Here, we make the first step  proving the existence
 of the ground state, which is
 a~$\Ga_d$-periodic stationary solution 
$\psi^0(\x)e^{-i\om ^0t}$,
$\phi^0(\x)$, 
$\ov\x=(\x^0_1,~\dots,\x^0_N)$  to the system 
(\re{LPS1})--(\re{LPS}): 
\beqn\la{LPS3} 
\h\om^0\psi^0(\x)&=&-\fr{\h^2}{2\cm}\De\psi^0(\x)+e\phi^0(\x)\psi^0(\x), 
~~~~\x\in T_d,
\\ 
\nonumber\\ 
-\De\phi^0(\x)&=&\rho^0(\x):= 
\si^0(\x)+e|\psi^0(\x)|^2, ~~~~~~\x\in T_d, 
\la{LPS4} 
\\ 
\nonumber\\ 
\la{LPS3g} 
0&=&- \5\langle\na\phi^0(\x), 
\mu^{\rm per}_j(\x-\x^0_j)\rangle, \qquad  j=1,\dots,N. 
\eeqn 
Here, $T_d:=\R^3/\Ga_d$ denotes  the `elementary  cell' of the crystal, 
$\langle\cdot,\cdot\rangle$ 
stands for the  scalar product in the Hilbert space $L^2(T_d)$ 
and its different extensions, and 
\be\la{rrr} 
\si^0(\x):= 
\sum_{j=1}^N\mu^{\rm per}_{j}(\x-\x^0_j),\quad \mu^{\rm per}_j(\x):= 
\sum_{\n\in\Z^d}\mu_j(\x-\x(\n)), 
\ee 
where we assume that the series converge in an appropriate sense. 
More precisely, we will construct a solution to the system (\re{LPS3})--(\re{LPS3g}) with $\si^0(\x)$ given by 
the first equation of (\re{rrr}) where  $\mu^{\rm per}_j$ satisfy the following condition:
\be\la{roL1}
\mbox{\bf Condition I.}\qquad\mu^{\rm per}_j\in L^1(T_d)\cap L^2(T_d),\qquad j=1,...,N.\qquad\qquad
\ee 
For 
instance, $\mu^{\rm per}_j\in  L^1(T_d)$ if $\mu_j\in  L^1(\R^3)$.
So we consider the case of smeared ions. The case of the point ions will be considered elsewhere.
In the cases $d=2$ and $d=1$ we will assume additional conditions  (\re{rous5}) and (\re{rous51})
respectively.

The elementary cell $T_d$ is isomorphic to the 3D torus for $d=3$, 
to the direct product of the 2D torus by $\R$ for $d=2$, 
and to the 
direct product of the 1D torus (circle)  by $\R^2$ for $d=1$. 

The system (\re{LPS3})--(\re{LPS3g}) is translation invariant similarly to (\re{LPS1})--(\re{LPS}).
Let us note that 
$\om^0$ should be real since
$\rIm\om^0\ne 0$ means an instability of the ground state:
the decay 
as $t\to\infty$
in the case $\rIm\om^0 < 0$ and the explosion  if $\rIm\om^0 > 0$.

Let us denote $\ds Z_j:=\int_{T_d}\mu^{\rm per}_j(\x)d\x/|e|$. Then 
\be\la{intro}
\int_{T_d}\si^0(\x)d\x=Z|e|,\qquad Z:=\sum_j Z_j. 
\ee 
The total charge per cell should be zero (cf. \ci{BBL2003}): 
\be\la{neu10} 
\int_{T_d} \rho^0(\x)d\x= 
\int_{T_d} [\si^0(\x) +e|\psi^0(\x)|^2]d\x=0. 
\ee 
For $d=3$  this neutrality condition
follows directly from equation (\re{LPS4}) by integration using  
$\Ga_3$-periodicity of $\phi^0(\x)$. For  $d=1$ and $d=2$ it follows from the finiteness
of energy per cell.
Equivalently,
the neutrality condition can be written as the normalization
\be\la{neuZ} 
\int_{T_d} |\psi^0(\x)|^2d\x 
=Z. 
\ee 
We allow arbitrary 
$Z_j\in\R$, however we
assume that  $Z>0$: otherwise the theory is trivial.

\smallskip

Let us comment on our approach. 
The neutrality condition (\re{neuZ}) 
defines the submanifold $\mathcal M$ in the space 
$H^1(T_d)\times T_d^N$
of 
space-periodic configurations $(\psi^0,\ov\x^0)$. 
We construct a~ground state as a~minimizer 
over $\mathcal M$
of the energy per cell 
(\re{HamsT}), (\re{HamsT22}), (\re{HamsT221}).

Our techniques in the 
case of 3D lattice is rather standard, and we use it 
as an `Ariadne's thread' to manage the more complicated 
cases of  2D and 1D lattices,  because the corresponding elementary cells 
are unbounded.

Namely, the derivation of the equations \eqref{LPS3}--\eqref{LPS3g} for the minimizer 
in the cases of  2D and 1D lattices
is not straightforward. The difficulty is that 
the energy per cell   is finite 
only 
on a dense subset of~$\mathcal M$ due to 
the infrared divergence of the Coulomb bond.
In these cases 
we restrict ourselves by one ion per cell, i.e., by $N=1$.
Then $\ov\x^0=\x^0_1$ can be chosen arbitrary
because of the translation invariance of the system  \eqref{LPS3}--\eqref{LPS3g}.
Respectively, 
now the energy per cell should be minimized over 
$\psi\in M$, where $M$ is the submanifold of $H^1(T_d)$
defined by the neutrality condition (\re{neuZ}).

\smallskip 
The  main  novelties 
of our approach  behind the 
technical proofs for 2D and 1D lattices are as follows:

\smallskip 
 
I. The energy per cell consists of two 
contributions: the kinetic energy, and the 
Coulomb bond. 
Generally, the Coulomb bond for  2D and 1D lattices
is infinite due to 
the infrared divergence
which 
is caused by the continuous spectrum of the Laplace operator
on the corresponding elementary cells.
The spectrum is continuous since 
the elementary cells are unbounded in the case of  
2D and 1D lattices in $R^3$. 
Let us note that the continuous spectrum and the infrared 
singularity also appear in the Schr\"odinger--\allowbreak 
Poisson molecular systems in $\R^3$ studied in 
\ci{Ben02,Kawohl2012,Nier93} where the singularity 
is summable, contrary to the space-periodic  case. 

We indicate suitable conditions (\re{rous5}), (\re{rous51}) which provide 
the finiteness of the Coulomb bond for a dense set of the fields in the case
of 2D and 1D lattice respectively.

Both contributions to the energy per cell (the kinetic energy and the 
Coulomb bond) are nonnegative.
Hence, 
for any minimizing sequence, both contributions 
are bounded. 
The bound for the kinetic energy ensures the compactness in 
each finite region of a~cell 
by the Sobolev embedding theorem.
However, 
this bound cannot prevent the decay of the electron field, i.e., 
its escape to infinity. 
Nevertheless, 
the Coulomb interaction prevents even the partial escape  to infinity,
as we show in Lemma \re{lM2}.
Physically this means that 
the electrostatic potential of the remaining positive charge 
becomes confining. 
 
\smallskip 
 
II. We construct the
solution  to the Poisson equation 
\eqref{LPS4}  
as the contour integral, 
providing the continuity and a bound for the electrostatic potential.
The main difficulty is a verification of the Schr\"odinger 
equation \eqref{LPS3} for the minimizer. Namely,
the Lagrange method of multipliers is not applicable
because
the energy per cell is infinite outside the submanifold 
$M\subset H^1(T_d)$ due to the infrared divergence
of the  Coulomb bond. 
Moreover, the Coulomb bond is infinite 
for a dense set of $\psi\in  M$. Hence, to differentiate 
the energy functional, we should construct the smooth paths in $ M$ 
lying outside this dense set.

\smallskip 
 
III. Finally, the proof that
$\om^0$ is real
(which is the stability condition for  the ground state)
is not straightforward
for  2D and 1D lattices, 
since the potential $\phi^0(x)$ a~priori can grow at infinity.
The correponding bounds for the potentials are given by 
\eqref{wfi2} and  \eqref{wfi1}.

\smallskip

The minimization strategy ensures the existence of a~ground state for any
 lattice~\eqref{Ga3}. One could expect
 that a stable lattice should provide a local minimum of
 the energy per  cell
 for fixed $d$, $N$ and  functions $\rho_j$, but this is still an open
problem.
% For example, the stability of a~1D lattice may result from  the
%dipole interaction providing a~local minimum of the energy per two
%neighboring cells.

\medskip 
 
Let us comment on related works. 
For atomic systems in $\R^3$, a~ground state was constructedby Lieb, Simon  and P.~Lions 
in the case of the Hartree and Hartree--\allowbreak Fock models
\ci{LS1977,Lions1981, Lions1987}, and 
by Nier
for the Schr\"odinger--\allowbreak Poisson  model \ci{Nier93}. 
The Hartree--\allowbreak Fock dynamics 
for molecular  systems in $\R^3$
has been constructed
by Canc\`es and Le Bris \ci{CB}.

A mathematical theory of the stability of  matter 
started from the pioneering works of 
Dyson, Lebowitz, Lenard, Lieb and others 
 for the Schr\"odinger many body model 
\ci{Dyson1967, Lieb2005, LL1972, Lieb2009}; 
see the survey in~\ci{Lemm}. 
Recently, the theory was extended to the 
quantized Maxwell field \ci{LL2005}.

These results and methods were developed 
last two decades
by Blanc, Le Bris, Catto, P. Lions and others 
to justify 
the  thermodynamic limit for the Thomas--Fermi and Hartree--\allowbreak Fock 
models 
with space-periodic  ion arrangement 
\ci{BBL2007,CBL1996,CBL1998,CBL2001} 
and to construct the corresponding space-periodic ground states 
\ci{CBL2002}, 
see the survey and further references in \ci{BL2005}. 
 
Recently, Giuliani, Lebowitz and Lieb have established 
the periodicity of the thermodynamic limit 
in 1D local mean field model 
without the assumption of periodicity of a~ion arrangement~\ci{GLL2007}. 
 
Canc\`es and others studied  short-range perturbations 
of the  Hartree--\allowbreak Fock 
model and proved that 
the 
density matrices of the perturbed and  unperturbed 
ground states 
differ by a~compact operator, 
\ci{CL2010,CS2012}. 
 
 %%%%%%%%%%%%%%%%%%%%%%%%%%%%%%%%%%%%%%%
The Hartree--Fock dynamics for infinite particle systems were considered recently 
by Cances and Stoltz \ci{CS2012}, and Lewin and Sabin \ci{LS2014-1}.
In \ci{CS2012}, the well-posedness  is established for 
local perturbations of the periodic ground state density matrix
in an infinite crystal.
However, the  space-periodic nuclear potential 
in the equation \ci[(3)]{CS2012}
is fixed that corresponds to 
the fixed nuclei positions. Thus the back reaction of the electrons onto 
the nuclei is neglected.
In \ci{LS2014-1},  the well-posedness is established for the 
von Neumann equation  with density matrices of infinite trace 
for pair-wise interaction potentials $w\in L^1(\R^3)$. Moreover, the authors  
prove the asymptotic stability of the ground state in 2D case \ci{LS2014-2}.
Nevertheless, the case of  the Coulomb potential for infinite particle systems remains open
since the corresponding generator is infinite.

%%%%%%%%%%%%%%%%%%%%%%%%%%%%%
 \medskip

Let us note that 2D and 1D crystals in $\R^3$ were not considered
previously. 
The space-periodic ground states for 1D and 2D lattices 
 give the model of
 the nanostructures similar to 
 the carbon nanotubes and graphene respectively.

\medskip 
 
The plan of our paper is as follows. 
In Section 2, we consider the $3$-dimensional lattice.
In Section 3, 
we construct a~ground state, derive equations 
\eqref{LPS3}--\eqref{LPS3g} and 
study smoothness properties of a~ground state 
for $2$-dimensional lattice. 
In  Section 4, we consider the  $1$-dimensional lattice. 
Finally, in Appendix we construct and estimate the potential for 1D lattice.

\medskip

{\bf Acknowledgments.} 
The author thanks H. Spohn for useful remarks and E. Kopylova for helpful discussions.

\setcounter{subsection}{0} 
\setcounter{theorem}{0} 
\setcounter{equation}{0} 
 
\section{3D lattice}

We consider the system  \eqref{LPS3}--(\re{LPS3g}) 
for the corresponding functions on the torus  $T_3=\R^3/\Ga_3$ 
and with $\x^0_j\5{\rm mod}\5\Ga_3\in T_3$. 
For $s\in\R$, we denote by $H^s$ the complex Sobolev 
space on the torus $T_3$, and for $1\le p\le \infty$, we denote by $L^p$ 
the complex Lebesgue space of  functions on $T_3$. 
 
\subsection{Energy per cell} 
The ground state will be constructed by minimizing the 
energy  in the cell $T_3$. 
To this aim, we will minimize the energy with respect to 
$\ov\x:=(\x_1,\dots,\x_N)\in(T_3)^N$ and 
$\psi\in H^1$ satisfying the neutrality condition 
(\re{neu10}): 
\be\la{neu1} 
\int_{T_3} \rho(\x)d\x=0,~~~~~~~~~~\rho(\x):=\si(\x) +e|\psi(\x)|^2.
\ee 
where we set 
\be\la{rrr2} 
\si(\x):= 
\sum_{j}\mu^{\rm per}_{j}(\x-\x_j), 
\ee 
similarly to (\re{rrr}). Let us note that $\rho\in L^2$ for $\psi\in H^1$
by our condition  (\re{roL1}) since $\psi\in L^6$ by the Sobolev embedding theorem.

We define the energy in the periodic cell 
for $\psi\in H^1$ by 
\beqn\la{HamsT} 
E(\psi,\ov\x)\!:=\!\! 
\int_{T_3}\Bigl[ 
\fr{\h^2}{2\cm}|\na\psi(\x)|^2 
+ 
\fr12\phi(\x)\rho(\x)\Bigr]d\x ,\qquad \phi(\x):=(-\De)^{-1}\rho, 
\eeqn 
where 
$(-\De)^{-1}\rho$ is well-defined by (\re{neu1}). 
Namely, consider the dual lattice 
\be\la{Ga3d} 
\Ga_3^*=\{\bk(\n)=\bb_1 n_1+\bb_2 n_2+\bb_3 n_3: \n=(n_1,n_2,n_3)\in\Z^3  \}, 
\ee 
where 
$\bb_k\bba_{k'}=2\pi\de_{kk'}$. 
Every function $\rho\in L^2$ admits the Fourier representation
\be\la{Fou} 
\rho(\x)=\fr1{\sqrt{|T_3|}}\sum_{\bk\in\Ga_3^*}\hat\rho(\bk) e^{-i\bk\x},\qquad 
\hat\rho(\bk)=\fr1{\sqrt{|T_3|}}\int e^{i\bk\x}\rho(\x)d\x.
\ee 
Respectively, we set 
\be\la{Fou2} 
\phi(\x)= 
(-\De)^{-1}\rho(\x) 
:= \fr1{\sqrt{|T_3|}}\sum_{\bk\in\Ga_3^*\setminus 0}\fr{\hat\rho(\bk)}{\bk^2} 
e^{-i\bk\x}. 
\ee 
This function 
$\phi\in H^2$
and satisfies 
the Poisson equation $-\De\phi=\rho$, 
since $\hat\rho(0)=0$ 
due to the neutrality condition (\re{neu1}).
Finally,
\be\la{Fou3} 
\int_{T_3}\phi(\x)d\x=0. 
\ee 
Now 
it is clear that the energy (\re{HamsT}) is finite for $\psi\in H^1$.
Let us rewrite the energy  as
\be\la{Upsi}
E(\psi,\ov\x)=I_1+I_2, 
\ee
where 
\beqn 
I_1(\psi)&:=&\fr{\h^2}{2\cm}\int_{T_3}|\na\psi(\x)|^2d\x\ge 0, 
\la{I1}\\ 
\nonumber\\ 
I_2(\phi)&:=& 
\fr12\int_{T_3} 
(-\De)^{-1}\rho(\x) 
\cdot\rho(\x) 
d\x=\fr12 \int_{T_3} 
|\na\phi(\x)|^2 
d\x\ge 0. \la{I2}
\eeqn 
The functional (\re{HamsT}) is chosen,  because  
\be\la{ener1} 
\fr{\de E}{\de\x_j}=
- \langle 
(-\De)^{-1}\rho(\x),\na\rho^{\rm per}_j(\x-\x_j)\rangle= 
 \langle\na\phi(\x),\rho^{\rm per}_j(\x-\x_j)\rangle, 
\ee
and the variational derivatives {\it formally} reads 
\be
\fr{\de E}{\de\Psi(\x)}= 
-2\fr{\h^2}{2\cm} 
\De\psi+ 
2 e 
(-\De)^{-1}\rho(\x)\psi(\x)= 
-2\fr{\h^2}{2\cm} 
\De\psi+ 
2e 
\phi(\x)\psi(\x).  \la{enerid} 
\ee
The variation in  (\re{enerid})  is taken 
over $\Psi(\x)=(\psi_1(\x),\psi_2(\x))\in L^2(T_3,\R^2)$, 
where 
$\psi_1(\x)=\rRe\psi(x)$ and $\psi_2(\x)=\rIm\psi(x)$. 
Respectively, all the terms in  (\re{enerid})  are identified with 
the corresponding 
$\R^2$-valued distributions.

\subsection{Compactness of minimizing sequence}

Our purpose here is to minimize the energy with respect to 
$$ 
(\psi,\ov\x)\in \mathcal M:=M\times T_3^N, 
$$ 
where $M$ denotes the manifold (cf. (\re{neuZ}))
\be\la{MZ} 
M=\{\psi\in H^1:~ \int_{T_3} |\psi(\x)|^2d\x 
=Z \}. 
\ee 
The energy is bounded from below since   
$
E(\psi,\ov\x)\ge 0 
$ 
by (\re{Upsi})-(\re{I2}).
We choose a minimizing sequence $(\psi_n, \ov\x_n)\in\mathcal M$ such that 
\be\la{min} 
E(\psi_n,\ov\x_n)\to E^0:=\inf_\mathcal M~ 
E(\psi,\ov\x), \qquad n\to\infty. 
\ee 
Our main result for a~3D lattice is the following: 
\bt\la{t3} 
Let condition \eqref{roL1} hold. Then
\medskip\\
i) There exists  $(\psi^0,\ov\x^0)\in \mathcal M$ with
\be\la{U0min} 
E(\psi^0,\ov\x^0)= E^0. 
\ee 
ii) Moreover,  $\psi^0\in H^2$ and satisfies equations \eqref{LPS3}--\eqref{LPS3g} with $d=3$,
where the  potential $\phi^0\in H^2$ is real, 
 and  $\om^0\in\R$. 

\et 
To prove item i),
let us denote 
\be\la{ron}
\rho_n(\x):=\si_n(\x)+e|\psi_n(\x)|^2,\qquad\qquad \si_n(\x):=\sum_{j}\mu^{\rm per}_{j}(\x-\x_{jn}).
\ee
Now the sequence $\psi_n$ and the corresponding sequence 
$\phi_n:=(-\De)^{-1}\rho_n$ 
are bounded in $H^1$ 
by (\re{Upsi})-(\re{I2}), (\re{Fou3}) and 
(\re{MZ})-(\re{min}). 
Hence, both sequences are precompact in~$L^p$ 
for any $p\in[1,6)$ by the Sobolev embedding theorem \ci{Adams,Sob}. 
Therefore, the sequence $\rho_n$ is precompact in $L^2$ by our assumption (\re{roL1}), and respectively, 
 the sequence $\phi_n$ is precompact in $H^2$.
As the result,
there exist a  subsequence $n'\to\infty$ for which  
\be\la{subs} 
\psi_{n'}\toLp\psi^0, \qquad 
\phi_{n'}\toHd\phi^0, \qquad
\ov\x_{n'}\to \ov\x^0 , \qquad \qquad n'\to\infty
\ee 
with any $p\in[1,6)$.
Respectively,
\be\la{subs1} 
\si_{n'}\toLd \si^0,  \qquad \rho_{n'}\toLd \rho^0, \qquad \qquad n'\to\infty,
\ee 
where $\si^0(\x)$ and $\rho^0(\x)$ are defined by (\re{rrr}) and (\re{LPS4}). 
Hence, the neutrality condition (\re{neu10}) holds, 
$(\psi^0,\ov\x^0)\in \mathcal M$, $\phi^0\in H^2$, and for these limit functions we have
\be\la{phi0}
-\De\phi^0=\rho^0, \qquad \ds\int_{{T_3}}\phi^0(\x)d\x=0.
\ee 
To prove identity (\re{U0min}), we take 
into account that 
$I_1(\psi)$ is lower semicontinuous on $L^2$, while
$I_2(\phi)$ is continuous on $H^2$; i.e., 
\be\la{sem} 
I_1(\psi^0)\le \liminf_{n'\to\infty} I_1(\psi_{n'}), 
\qquad I_2(\phi^0)= \lim_{n'\to\infty} I_2(\phi_{n'}). 
\ee 
These limits, together with (\re{min}),  imply that 
\be\la{sem2} 
E(\psi^0,\ov\x^0) 
=I_1(\psi^0)+I_2(\phi^0) 
\le E^0. 
\ee 
Now (\re{U0min}) follows from the definition of $E^0$, since 
$(\psi^0,\ov\x^0)\in\mathcal M$. Thus Theorem \re{t3} i) is proved.
\medskip

We will prove the item ii) in next sections.

\subsection{Variation of the energy} 

Theorem \re{t3} ii) follows from next proposition.

\bp\la{tgs23} The limit functions 
\eqref{subs}
satisfy equations {\rm \eqref{LPS3}--\eqref{LPS3g}} with $d=3$
and $\om^0\in\R$.

\ep

Equation (\re{LPS4}) is proved in (\re{phi0}), and the equation
 \eqref{LPS3g} follows from (\re{ener1}) and (\re{U0min}). 
 It remains to prove the Schr\"odinger equation \eqref{LPS3}. 
Let us denote $\E(\psi):=E(\psi,\ov\x^0)$.
We derive \eqref{LPS3} in next sections, 
equating the variation of $\E(\cdot)|_{M}$ to zero at $\psi=\psi^0$. 
In this section we calculate the corresponding G\^ateaux variational derivative.

We should work directly on~$M$ introducing an atlas in a~neighborhood 
of $\psi^0$ in $M$.
We define the atlas 
as the stereographic projection from the tangent plane 
$TM(\psi^0)=(\psi^0)^\bot:=\{\psi\in H^1: 
\langle\psi,\psi^0\rangle=0\}$ 
to the sphere (\re{MZ}): 
\be\la{3atlas} 
\psi_\tau= \fr{\psi^0+\tau~~~}{\Vert\psi^0+\tau\Vert_{L^2}} 
\sqrt{Z}, \qquad  \tau\in (\psi^0)^\bot. 
\ee 
Obviously, 
\be\la{3tau} 
\fr d{d\ve}\Big|_{\ve=0} \psi_{\ve\tau}=\tau,\qquad \tau\in (\psi^0)^\bot, 
\ee 
where the derivative exists in $H^1$. We define
the `G\^ateaux derivative' of $\E(\cdot)|_{M}$  as 
\be\la{3Gder} 
D_\tau \E(\psi^0):=\lim_{\ve\to 0}\fr{\E(\psi_{\ve\tau})-\E(\psi^0)}{\ve},
\ee 
if this limit exists. We should restrict the set of allowed tangent vectors~$\tau$. 
 
\bd 
$\ccT^0$ is the space of test functions 
$\tau\in(\psi^0)^\bot\cap C^\infty(T_3)$. 
\ed 
Obviously, $\ccT^0$ is dense in $(\psi^0)^\bot$ in the norm of  $H^1$. 
 Let us rewrite the energy (\re{HamsT}) as
 \be\la{3HamsT22} 
\E(\psi):= 
\int_{T_3}\Bigl[ 
\fr{\h^2}{2\cm}|\na\psi(\x)|^2 
+ 
\fr12 |\Lam\rho(\x)|^2 
\Bigr]d\x, \qquad \rho(\x):=\si(\x) +e|\psi(\x)|^2, 
\ee 
 where $\Lam:=(-\De)^{-1/2}$ is defined similarly to (\re{Fou2}):
 \be\la{Fou2L} 
\Lam\rho(\x) 
:= \fr1{\sqrt{|T_3|}}\sum_{\bk\in\Ga_3^*\setminus 0}\fr{\hat\rho(\bk)}{|\bk|} 
e^{-i\bk\x}\in L^2 \qquad \mbox{\rm for } \qquad  \rho\in L^2. 
\ee

\bl\la{3lvar} 
Let 
$\tau\in\ccT^0$. Then the  derivative \eqref{3Gder} exists, and  {\rm (}cf. \eqref{enerid}{\rm )}, 
\be\la{3Gder2} 
D_\tau \E(\psi^0)=\int_{T_3}  \Big[\fr{\h^2}{2\cm} 
(\na\tau\ov{\na\psi^0}+\na\psi^0\ov{\na \tau}) 
+e {\Lam\rho^0}\Lam(\tau\ov{\psi^0}+\psi^0\ov \tau )\Big]d\x. 
\ee 
\el 
\Pr 
Let us denote $\rho_{\ve\tau}(\x):=\si^0(\x)+e|\psi_{\ve\tau}(\x)|^2$.

\bl\la{3lL2}
 For $\tau\in\ccT^0$ we have
 \be\la{3Gder3} 
D_\tau \Lam\rho:= 
\lim_{\ve\to 0}\fr{\Lam\rho_{\ve\tau}-\Lam\rho^0}{\ve} 
 =e\Lam(\tau\ov{\psi^0}+\psi^0\ov \tau ), 
\ee 
where the limit converges in  $L^2$.

\el
\Pr In the polar coordinates
\be\la{3al} 
\psi_{\ve\tau}=(\psi^0+\ve\tau)\cos\al,\qquad 
\al=\al(\ve)=\arctan\fr{\ve\Vert\tau\Vert_{L^2}}{\Vert\psi^0\Vert_{L^2}}. 
\ee 
Hence,
\beqn\la{3rod} 
\Lam\rho_{\ve\tau}&=&\Lam\si^0 +e\cos^2\al\Lam|\psi^0+\ve\tau|^2 
\nonumber\\
\nonumber\smallskip\\
&=& 
\Lam\rho^0+e\ve\cos^2\al\Lam(\tau\ov{\psi^0}+\psi^0\ov\tau)+ 
e\Lam[\ve^2|\tau|^2\cos^2\al-|\psi^0|^2\sin^2\al]. 
\eeqn 
Here 
$\Lam\rho^0\in L^2$ since 
$\rho^0\in L^2$, and similarly $\Lam [\psi^0\ov\tau]\in L^2$ since 
$\psi^0 \ov \tau\in L^2$.
It remains to 
estimate 
 the last term 
of (\re{3rod}),
\be\la{3lt}
R_\ve:=\Lam[\ve^2|\tau|^2\cos^2\al-|\psi^0|^2\sin^2\al].
\ee
Here $|\psi^0|^2\in L^2$ 
since $\psi^0\in H^1\subset L^6$. Finally,  $|\tau|^2\in L^2$ and $\sin^2\al\sim \ve^2$.
Hence, the convergence (\re{3Gder3}) holds in $L^2$.\bo
\medskip

Now (\re{3Gder2}) follows by differentiation 
in $\ve$ of (\re{3HamsT22}) with $\psi=\psi_{\ve\tau}$ and $\rho=\rho_{\ve\tau}$.
\bo

 \subsection{The variational identity} 
 
Since $\psi^0$ is a minimal point, the G\^ateaux derivative (\re{3Gder2}) 
vanishes: 
\be\la{3GaD} 
\int_{T_2}  \Big[\fr{\h^2}{2\cm} 
(\na\tau\ov{\na\psi^0}+\na\psi^0\ov{\na \tau}) 
+
e {\Lam\rho^0}\Lam(\tau\ov{\psi^0}+\psi^0\ov \tau )
\Big]d\x=0. 
\ee 
Substituting $i\tau$ instead of $\tau$ in this identity 
and subtracting, we obtain 
\be\la{3GaD2} 
-\fr{\h^2}{2\cm}\langle\De\psi^0, \tau\rangle 
+ e \langle 
\Lam\rho^0,\Lam(\ov{\psi^0} \tau) \rangle=0. 
\ee 
Next step 
we should evaluate the ``nonlinear'' term.

\bl For the limit functions 
\eqref{subs}--\eqref{subs1} we have 
\be\la{3sp} 
\langle 
\Lam\rho^0,\Lam(\ov{\psi^0} \tau) \rangle= 
\langle\phi^0\psi^0, \tau \rangle,\qquad \tau\in\ccT^0.
\ee

\el 
\Pr 
Let us substitute 
$\rho^0=-\De\phi^0$. 
 Then, by the Parseval--\allowbreak Plancherel identity, 
\be\la{3GaD3} 
\langle 
\Lam\rho^0,\Lam(\ov{\psi^0} \tau) \rangle= \sum_{\bk\in\Ga_3^*\setminus 0}
\fr{\bk^2\hat\phi^0(\bk)}{|\bk|}\cdot
\fr{\widehat{\ov{\psi^0} \tau}(\bk)}{|\bk|} = 
\langle\hat \phi^0,\widehat{\ov{\psi^0} \tau} \rangle
=\langle \phi^0,\ov{\psi^0} \tau \rangle=\langle \phi^0\psi^0, \tau \rangle. 
\ee 
which proves (\re{3sp}).\bo 
\medskip 
 
Using (\re{3sp}), we can rewrite (\re{3GaD2}) as the variational identity (cf.\ (\re{enerid})) 
\be\la{3GaD22} 
\langle -\fr{\h^2}{2\cm}\De\psi^0 
+ e 
\phi^0\psi^0,\tau \rangle=0,    \qquad \tau\in\ccT^0. 
\ee 
 
\subsection{The Schr\"odinger equation} 
 
Now we prove the Schr\"odinger equation (\re{LPS3}) with $d=3$.
\bl\la{lse}
$\psi^0$ is the eigenfunction of the 
Schr\"odinger operator $H=-\fr{\h^2}{2\cm}\De+e\phi^0$:
\be\la{3Hpsi} 
H\psi^0=\lam\psi^0,
\ee 
where 
$\lam\in\R$. 
\el
\Pr
First, $H\psi^0$ is a well-defined distribution since $\phi^0\in H^2\subset C(T_3)$ by (\re{subs}).
Second,
 $\psi^0\ne 0$ since $\psi^0\in M$ and $Z>0$. Hence, there exists a~test function 
$\theta\in C^\infty(T_3)\setminus\ccT^0$, i.e., 
\be\la{3test} 
\langle\psi^0,\theta\rangle\ne 0. 
\ee 
Then 
\be\la{3test2} 
\langle(H-\lam)\psi^0,\theta\rangle= 0. 
\ee 
for an appropriate $\lam\in\C$. However, 
$(H-\lam)\psi^0$ also  annihilates $\ccT^0$ by (\re{3GaD22}), 
hence it  annihilates the whole space 
$C^\infty(T_3)$. This implies (\re{3Hpsi}) in the sense of 
distributions with a $\lam\in\C$. Finally,
the potential is  real, and 
 $\phi^0\in C(T_3)$. Hence,
 $\lam\in\R$.\bo
 \medskip

  This lemma implies  equation (\re{LPS3})  with $\hbar\om^0=\lam$.
   Hence, $\psi^0\in H^2$ since $\phi^0\in C(T_3)$.
Now Theorem \re{t3} ii) is proved.

\subsection{Smoothness of  ground state} 

We have proved that $\psi^0\in H^2$ under condition (\re{roL1}).
Using the Schr\"odinger equation (\re{3Hpsi})
we can improve further the smoothness
 of $\psi^0$ strengthening  the condition (\re{roL1}).
Namely, let us assume that
\be\la{rpem3} 
\mu_j^{\rm per}\in C^\infty(T_3), \qquad j=1,...,N.
\ee 
Then also
\be\la{rper} 
\si^0(\x):=\sum_{j=1}^N \mu^{\rm per}_j(\x-\x^0_j)\in C^\infty(T_3). 
\ee 
For example, (\re{rpem3}) and (\re{rper}) hold if $\mu_j\in  \cS(\R^3)$,
where $\cS(\R^3)$ is the Schwartz space of test functions.

\bl\la{lpp} 
Let  condition \eqref{rpem3} hold, and 
$\psi^0\in H^2$,  $\phi^0\in H^2$ be a solution to equations 
\eqref{LPS3}--\eqref{LPS3g} with $d=3$ and some $\x\in T_3^N$. Then 
the functions $\psi^0$ and $\phi^0$ are smooth. 
\el 
\Pr 
First, $\phi^0\psi^0\in H^2$
since  $H^s$ is the algebra for $s>3/2$. 
Hence, equation ~\eqref{LPS3} implies that 
\be\la{HC2} 
\psi^0\in H^4\subset C^2(T_3). 
\ee 
Now 
$\rho^0:=\si^0+e|\psi^0|^2\in H^4$ by \eqref{rpem3}. Then~\eqref{LPS4} implies that $\phi^0\in H^6\subset C^4(T_3)$.
Hence, $\phi^0\psi^0\in H^4$, $\psi^0\in H^6$, $\rho^0\in H^6$, etc.\bo

\setcounter{equation}{0} 
 
\section{2D lattice} 
 
For simplicity of notation we will consider the 2D lattice $\Ga_2=\Z^2$
 and 
construct a solution to system   \eqref{LPS3}--\eqref{LPS3g} 
for the corresponding functions on the `cylindrical cell' 
 $T_2:=\R^3/\Ga_2=\T^2\times\R$ with  
 the coordinates $\x=(x_1,x_2,x_3)$, where $(x_1,x_2)\in \T^2$
and $x_3\in\R$. 
Now we denote by $H^s$ the complex Sobolev space on $T_2$, and 
by $L^p$, the  complex Lebesgue space of functions on $T_2$.

We will construct a~ground state by minimizing the  
energy  (\re{HamsT}), 
where the integral is extended 
over $T_2$ instead of $T_3$. 
The neutrality condition 
of type (\re{neu1})
holds for 
$\Ga_2$-periodic states with finite energy, as we show below.

\subsection{The energy per cell} 
We restrict ourselves by $N=1$, so $\ov\x^0=\x^0_1$ can be chosen arbitrary
because of the translation invariance of the system  \eqref{LPS3}--\eqref{LPS3g}.
For example, we can set $\x^0_1=0$.

The  energy in the cylindrical cell $T_2$
is defined similarly to  (\re{HamsT}), which we rewrite as  (\re{3HamsT22}):
\be\la{HamsT22} 
\E(\psi):= 
\int_{T_2}\Bigl[ 
\fr{\h^2}{2\cm}|\na\psi(\x)|^2 
+ 
\fr12 |\Lam\rho(\x)|^2 
\Bigr]d\x, \qquad \rho(\x):=\si^0(\x) +e|\psi(\x)|^2. 
\ee 
Here $\si^0(\x)$ is defined by 
 (\re{rrr2}) with $N=1$ and $\x^0_1=0$:
\be\la{rrr22} 
\si^0= 
\mu^{\rm per}_1\in L^1\cap L^2
\ee 
according to our condition (\re{roL1}).
Hence, we have 
\be\la{intro2} 
\ds\int_{T_2} 
\si^0(\x)d\x=Z_1|e|,~~~~~Z_1>0. 
\ee  
Further, $\Lam$ is the operator $(-\De)^{-1/2}$ defined by 
the Fourier transform. Namely, we denote
 $\Ga_2^*=2\pi\Ga_2$, 
and define the Fourier representation for the test functions 
$\vp\in C_0^\infty(T_2)$ by 
\be\la{FR2} 
\vp(\x)= 
\fr 1{\sqrt{2\pi}} \sum_{\bk\in \Ga_2^*} e^{-i(\bk_1x_1+\bk_2x_2)} 
\int_\R e^{-i\xi x_3}\hat \vp(\bk,\xi)d\xi, \qquad \x\in T_2,
\ee 
where 
\be 
\la{FT2} 
\hat 
\vp(\bk,\xi)=F\vp(\bk,\xi)=\fr 1{\sqrt{2\pi}} 
\int_{T_2} e^{i(\bk_1x_1+\bk_2x_2+\xi x_3)} \vp(\x)d\x,
\qquad  (\bk,\xi)\in\Si_2:=\Ga_2^*\times \R.
\ee 
The operator $\Lam$ is defined for $\vp\in L^1\cap L^2$ by 
\be\la{Lam} 
\Lam \vp=F^{-1} \fr{\hat \vp(\bk,\xi)}{\sqrt{\bk^2+\xi^2}} 
\ee 
provided the quotient belongs to $L^2(\Si_2)$. In this case
\be\la{02} 
\hat \vp(0,0)=0. 
\ee 
 Let us note that $\rho\in L^1\cap L^2$ for $\psi\in H^1$
by our condition  (\re{roL1}) since $\psi\in L^p$ with 
$p\in [2,6]$ 
by the Sobolev embedding theorem.
For $\psi\in H^1$ 
with finite energy (\re{HamsT22}) 
we have $\Lam\rho\in L^2(\Si_2)$. Therefore,
(\re{02}) with  $\vp=\rho$ implies 
the neutrality condition 
(\re{neu1}) with $T_2$ instead of $T_3$: 
\be\la{neu2} 
\hat\rho(0,0)=\int_{T_2} \rho(\x)d\x= 
\int_{T_2} [\si^0(\x) +e|\psi(\x)|^2]d\x=0. 
\ee 
Now (\re{intro2}) gives
\be\la{M2} 
\int_{T_2}
|\psi(\x)|^2d\x=Z_1.
\ee 
In other words, the finiteness of the Coulomb energy 
$\Vert\Lam\rho\Vert^2$ prevents the electron charge from escaping to infinity, 
as mentioned in Introduction. 
 \bd 
$M_2$ denotes the set of  
$\psi\in H^1$ satisfying the neutrality condition \eqref{M2}. 
\ed 
It is important that the energy be finite for a nonempty set of $\psi\in H^1$.
To find the corresponding condition, let us rewrite the 
energy (\re{HamsT22})
using the Parseval-Plancherel identity:
\be\la{HamsT22F} 
\E(\psi)= 
\sum 
_{\bk\in \Ga_2^*} 
\fr{\h^2}{2\cm}
\int_\R(\bk^2+\xi^2)|\hat\psi(\bk,\xi)|^2 d\xi
+ 
\fr12 
\sum 
_{\bk\in \Ga_2^*} 
\int_\R
\fr {|\hat\rho(\bk,\xi)|^2} {\bk^2+\xi^2} d\xi. 
\ee 
 Here the first term on the right hand side is finite for all $\psi\in H^1$.
The second term is finite up to the infrared divergence at 
the point $(\bk,\xi)=(0,0)$
since $\rho\in L^2(\Si_2)$ for $\psi\in H^1$.

 We note that (\re{intro2}) can be written as
$\hat\mu^{\rm per}_1(0)+eZ_1=0$. 
We will assume that moreover,
\be\la{rous5} 
\mbox{\bf Condition II.}\qquad
\fr{\hat\mu^{\rm per}_1(0,\xi)+eZ_1}{|\xi|}\in L^2(-1,1). \qquad\qquad\qquad\qquad
\ee 
For example, this condition holds, 
%by Sobolev's embedding theorem \ci{LM,Mazia,Sob}
provided that  
\be\la{rous5if} 
\int_{T_2}  |x^3||\mu^{\rm per}_1(\x)|d\x <\infty.
\ee 

\bl\la{lne}
Let conditions \eqref{roL1} and \eqref{rous5} hold, $N=1$ and $\x^0_1\in T_2$. Then 
the energy \eqref{HamsT22F} is finite for a dense set of   $\psi\in H^1$.

\el
\Pr By definition, $\hat \rho(0,\xi)=\hat\mu^{\rm per}_1(0,\xi)+e\hat P(0,\xi)$, where $P(\x):=|\psi(\x)|^2$.
Hence, (\re{rous5}) implies that
the energy (\re{HamsT22F}) is finite for   $\psi\in M_2$ with finite momenta 
$
\ds\int_{T_2}  |x^3|~|\psi(\x)|^2d\x<\infty.
$
\bo

\medskip 
\subsection{Compactness of minimizing sequence} 
 
Similarly to the 3D case, the energy is nonnegative, and 
we choose a minimizing sequence $\psi_n\in  M_2$ 
such that 
\be\la{min2} 
\E(\psi_n)\to \E^0:=\inf_{M_2}~ 
\E(\psi),\qquad n\to\infty. 
\ee 

The second main result of the present paper is the following.
 
\bt\la{tgs2} 
Let conditions \eqref{roL1} and \eqref{rous5} hold, and $N=1$. Then 
\medskip\\
i) There exists $\psi^0\in M_2$ with
\be\la{U0min2} 
\E(\psi^0)= \E^0. 
\ee 
ii) Moreover,  $\psi^0\in H^2_{\rm loc}(T_2)$ and  satisfies equations \eqref{LPS3}--\eqref{LPS3g} with $d=2$,
where 
the potential  $\phi^0\in H^2_{\rm loc}(T_2)$ is real, $\x^0_1=0$,
 and  $\om^0\in\R$. 
\medskip\\
iii) The following bound holds
\be\la{wfi2} 
|\phi^0(\x)|\le 
C(1+|x_3|)^{1/2},\qquad\qquad \x\in T_2.
\ee

\et
To prove item i),
let us note 
that the sequence $\psi_n$  
is bounded in $H^1$
due to 
(\re{HamsT22}),
(\re{M2}) and (\re{min2}).
Hence,  by the Sobolev embedding theorem \ci{Adams,Sob}, 
the sequence $\psi_n$  
is bounded in $L^p$ with each $p\in[2,6)$ and
compact in $L^p_R:=L^p(T_2(R))$ 
for  any $R>0$, 
where $T_2(R)=\{\x\in T_2:~|x_3|<R\}$. 
Therefore, there exists a~subsequence 
\be\la{subs2} 
\psi_{n'}\toLpR\psi^0,\qquad \rho_{n'}:=\mu^{\rm per}_1+e|\psi_{n'}|^2\toLLR\rho^0,
\qquad 
n'\to\infty,\quad \forall R>0,
\ee 
since 
$\mu^{\rm per}_1\in L^1\cap L^2$ by (\re{roL1}).
Hence, $\psi^0\in H^1\cap L^p$,  and
\be 
\la{pr} 
\rho^0(\x)=\mu^{\rm per}_1(\x)+e|\psi^0(\x)|^2\in L^1\cap L^2. 
\ee 
Next problem is to  check the neutrality condition (\re{M2}) for the 
limit charge density $\rho^0$ since the convergence (\re{subs2}) itself is not sufficient.

\bl\la{lM2} The limit function
$\psi^0\in M_2$, 
and the energy \eqref{HamsT22} for $\psi^0$ is finite. 
\el
\Pr
Let us prove that
\be\la{UU} 
\E(\psi^0)\le \E^0. 
\ee 
Indeed, 
(\re{HamsT22F}) with $\psi=\psi_{n'}$  reads 
\be \la{En}
\E(\psi_{n'}):= 
\Bigl\langle 
\fr{\h^2}{2\cm}|f_{n'}(\bk,\xi)|^2 
+ 
\fr 12 |g_{n'}(\bk,\xi)|^2 
\Bigr\rangle_{\Si_2}, 
\ee 
where $\langle \dots \rangle_{\Si_2}$ stands for 
$\ds\sum_{\bk\in \Ga_2^*}\int_\R~\dots~d\xi$ and 
$$ 
f_{n'}(\bk,\xi):=\sqrt{\bk^2+\xi^2} 
\hat\psi_{n'}(\bk,\xi),\qquad g_{n'}(\bk,\xi):=\fr {\hat\rho_{n'}(\bk,\xi)} 
{\sqrt{\bk^2+\xi^2}}. 
$$ 
The  functions  $\hat\psi_{n'}$ and $\hat\rho_{n'}$
are bounded in $L^2(\Si_2)$, and are 
converging in the sense of distributions due to (\re{subs2}). Hence,
\be\la{tow2} 
\hat\psi_{n'}\tow \hat\psi^0,\qquad\hat\rho_{n'}\tow \hat\rho^0,
\qquad  n'\to\infty.
\ee
Similarly, the  functions $f_{n'}$ and $g_{n'}$ 
are bounded in $L^2(\Si_2)$ by (\re{En}), (\re{min2}), and are
converging in the sense of distributions due to (\re{tow2}).
Therefore, 
\be\la{tow} 
f_{n'}\tow f^0,\qquad g_{n'}\tow g^0,\qquad n'\to\infty. 
\ee 
Hence, for the limit functions, 
$$ 
f^0(\bk,\xi)=\sqrt{\bk^2+\xi^2} 
\hat\psi^0(\bk,\xi), 
\qquad g^0(\bk,\xi)=\fr {\hat\rho^0(\bk,\xi)} 
{\sqrt{\bk^2+\xi^2}}, \qquad\qquad  a.a. \,\,\, (\bk,\xi)\in\Si_2.
$$ 
Therefore, (\re{UU}) holds since  
\be\la{Uro} 
\E(\psi^0)= 
\Bigl\langle 
\fr{\h^2}{2\cm}|f^0(\bk,\xi)|^2 
+ 
\fr 12 |g^0(\bk,\xi)|^2 
\Bigr\rangle_{\Si_2}\le \E^0 
\ee 
by the week convergence (\re{tow}). 
In particular,  
\be\la{Lamro}
\Lam\rho^0\in L^2.
\ee
Therefore, $\hat\rho^0(0,0)=0$ as in (\re{neu2}) since $\rho^0\in L^1$ by (\re{pr}).
Hence,  $\psi^0\in  M_2$.\bo
 \medskip\\
Now (\re{UU}) implies 
(\re{U0min2}). Thus Theorem \re{tgs2} i) is proved.

\subsection{The Poisson equation} 
Our aim here is to construct the potential which is the solution to 
the Poisson equation 
(\re{LPS4}) with $d=2$. It suffices to solve the equation 
\be\la{rho2} 
\na\phi^0(\x)=G^0(\x),\qquad\qquad \x\in T_2,
\ee
where 
$G^0(\x):= 
- 
 iF^{-1}\fr{(\bk,\xi)}{\bk^2+\xi^2}\hat\rho^0(\bk,\xi)$ 
  is a~real vector field, 
$G^0\in L^2\otimes \R^3$ by (\re{Lamro}), and $\rot G^0(\x)\equiv 0$.

\bl\la{lP2}
The equation \eqref{rho2} 
admits real  solution $\phi^0\in H^2_{\rm loc}(T_2)$
which is unique up to an additive constant, and satisfies the bound \eqref{wfi2}.

\el
\Pr  The uniqueness up to constant is obvious. If the solution exists, then 
$\phi^0\in H^2_{\rm loc}(T_2)$  by (\re{pr}). Local solutions exist since  $\rot G^0(\x)\equiv 0$. However, the 
existence of the global solution is not obvious since the cell $T_2$ is not 1-connected.

We will prove the existence using the following arguments. 
Formally $\phi^0(x)=F^{-1}\fr{\hat\rho^0(\bk,\xi)}{\bk^2+\xi^2}$. However, the last expression is not correctly defined distribution
in the neighborhood of the point $(0,0)$. To avoid this infrared divergence, we split $\hat\rho^0=\hat\rho_1+\hat\rho_2$ where 
\be\la{split}
 \hat \rho_1(\bk,\xi)=\left\{
 \ba{ll}
 \hat \rho^0(0,\xi),& \bk=0,~|\xi|<1,\\
 0,& \mbox{\rm otherwise.}
 \ea\right.
\ee
Respectively, 
$G^0=G_1+G_2$, and
the solution $\phi^0=\phi_1+\phi_2$. Obviously,
\be\la{G1}
G_1(\x)=- 
 iF^{-1}\fr{(0,\xi)}{\xi^2}\hat\rho_1(0,\xi)=\e_3g_1(x_3),\qquad \e_3:=(0,0,1),
\ee
and $g_1(x_3)$ is a smooth function. Moreover, 
(\re{pr}) implies that $g_1(x_3)$
 is the real function, and $g_1\in L^2(\R)$ since $G^0\in L^2\otimes \R^3$.
Hence, the solution $\phi_1(\x)=\ds\int_0^{x_3} g_1(s)ds$ is smooth and  continuous, 
and depends on $x_3$ only.
The bound (\re{wfi2}) for $\phi_1$ follows by the Cauchy-Schwartz inequality.

The second solution is given by 
$\phi_2(\x)=F^{-1}\fr{\hat\rho_2(\bk,\xi)}{\bk^2+\xi^2}$, where  $\hat\rho_2\in L^2(\Si_2)$ by (\re{pr}). Moreover, $\hat\rho_2(0,\xi)=0$ for $|\xi|<1$, and hence
$\phi_2\in H^2$.\bo
\medskip
\brs
i) The function $\phi^0(\x)=(1+|x_3|)^{1/2-\ve}$ with $\ve>0$ 
shows that the bound \eqref{wfi2} is exact under the condition $\na\phi^0\in L^2$. 
Note that the potential of uniformly charged plane grows linearly with the distance. 
\medskip\\
ii)
In the Fourier transform, \eqref{rho2} implies that 
\be\la{fm2} 
(\bk,\xi)\hat\phi^0(\bk,\xi)\in L^2(\Si_2)\otimes\C^3. 
\ee 
 \ers

\subsection{Variation of the energy} 

Theorem \re{tgs2} ii) follows from next proposition.

 \bp\la{tgs22} 
The functions $\psi^0$, $\phi^0$ 
satisfy equations {\rm \eqref{LPS3}--\eqref{LPS3g}} with $d=2$
and $\om^0\in\R$.

\ep

The equation (\re{LPS4}) is
 proved above, and the equation
 \eqref{LPS3g} follows from (\re{ener1}) and (\re{U0min2})
 by the translation invariance of the energy.
 It remains to prove the Schr\"odinger equation \eqref{LPS3}. 
We are going to derive \eqref{LPS3}, 
equating the variation of $\E(\psi)|_{M_2}$ to zero at $\psi=\psi^0$. 
 In this section we calculate the corresponding G\^ateaux variational derivative.

 Similarly to (\re{3atlas}),
we define the atlas in a~neighborhood 
of $\psi^0$ in $M_2$ 
as the stereographic projection from the tangent plane 
$TM_2(\psi^0)=(\psi^0)^\bot:=\{\psi\in H^1: 
\langle\psi,\psi^0\rangle=0\}$ 
to the sphere (\re{M2}): 
\be\la{atlas} 
\psi_\tau= \fr{\psi^0+\tau~~~}{\Vert\psi^0+\tau\Vert_{L^2}} 
\sqrt{Z_1}, \qquad  \tau\in (\psi^0)^\bot. 
\ee

\bd 
$\ccT^0$ is the space of test functions 
$\tau\in(\psi^0)^\bot\cap C_0^\infty(T_2)$. 
\ed 
 
Obviously, $\ccT^0$ is dense in $(\psi^0)^\bot$ in the norm of  $H^1$. 
 
\bl\la{lvar} 
Let $\tau\in\ccT^0$. Then 
 \medskip\\
{\rm i)} 
The energy 
$\E(\psi_{\ve\tau})$ is finite for $\ve\in\R$. 
 \medskip\\
{\rm ii)} The  G\^ateaux derivative \eqref{3Gder} exists, and  similarly to  \eqref{3Gder2},
\be\la{Gder2} 
D_\tau \E(\psi^0)=\int_{T_2}  \Big[\fr{\h^2}{2\cm} 
(\na\tau\ov{\na\psi^0}+\na\psi^0\ov{\na \tau}) 
+e {\Lam\rho^0}\Lam(\tau\ov{\psi^0}+\psi^0\ov \tau )\Big]d\x. 
\ee 
\el 
\Pr 
i)  We should prove the bound 
\be\la{HamsT223} 
\E(\psi_{\ve\tau}):= 
\fr{\h^2}{2\cm}\int_{T_2}|\na\psi_{\ve\tau}(\x)|^2d\x 
+ 
\fr12 \int_{T_2}|\Lam\rho_{\ve\tau}(\x)|^2 
d\x<\infty, 
\ee 
where $\rho_{\ve\tau}(\x):=\si^0(\x)+e|\psi_{\ve\tau}(\x)|^2$.
The first integral in (\re{HamsT223}) 
is finite, since $\psi_{\ve\tau}\in H^1$. 

\bl\la{lL2}
 $\Lam\rho_{\ve\tau}\in L^2$ for $\tau\in\ccT^0$ and  $\ve\in\R$, and
 \be\la{Gder3} 
D_\tau \Lam\rho:= 
\lim_{\ve\to 0}\fr{\Lam\rho_{\ve\tau}-\Lam\rho^0}{\ve} 
 =e\Lam(\tau\ov{\psi^0}+\psi^0\ov \tau ), 
\ee 
where the limit converges in  $L^2$.
\el
\Pr 
We use the polar coordinates  (\re{3al}) and the corresponding representation
(\re{3rod}):
\be\la{3rod2} 
\Lam\rho_{\ve\tau}= 
\Lam\rho^0+e\ve\cos^2\al\Lam(\tau\ov{\psi^0}+\psi^0\ov\tau)+ 
e\Lam[\ve^2|\tau|^2\cos^2\al-|\psi^0|^2\sin^2\al]. 
\ee 
Now $\Lam\rho^0\in L^2$ according to (\re{Lamro}). 
Further, $\Lam [\tau\ov{\psi^0}]\in L^2$ by the following arguments:
\medskip\\
a) $\tau\ov{\psi^0}  \in L^2$, 
\medskip\\
b) $\widehat{\tau\ov{\psi^0}}$ is the smooth function on $\Si_2$,
and 
\medskip\\
c) the orthogonality $\tau\bot\psi^0$ implies
that
\be\la{T0} 
\widehat{\tau\ov{\psi^0}}(0,0)=0. 
\ee 
It remains to estimate 
 the last term 
of (\re{3rod2}),
Let us denote $T(\x):=|\tau(\x)|^2$ and  $P(\x):=|\psi^0(\x)|^2$. Then the last  
term (up to a constant factor) reads
\be\la{lt}
R_\ve(\x):=\Lam[\ve^2 T(\x)
\cos^2\al-P(\x)\sin^2\al].
\ee
\bl\la{l37}
$R_\ve\in L^2$
for $\ve\in\R$, and
\be\la{ltm} 
\Vert R_\ve\Vert_{L^2}=\cO(\ve^2),~~~~~~~\ve\to 0.
\ee
\el
\Pr
{\it i)} It suffices to check that 
\beqn \la{sc} 
&&\fr{\ve^2{\hat T(0,\xi)}\cos^2\al-{\hat P(0,\xi)}\sin^2\al} 
{|\xi|} 
\nonumber\\ 
\nonumber\\ 
&=& 
\fr{(\ve^2{\hat T(0,\xi)}-Z_1\tan^2\al)\cos^2\al}{|\xi|}- 
\fr{({\hat P(0,\xi)}-Z_1)\sin^2\al}{|\xi|} 
\in L^2(-1,1). 
\eeqn 
Let us consider each term of the last line of (\re{sc}) separately.
\medskip

1) The first  quotient
belongs to  $L^2(-1,1)$, since 
\be\la{las3} 
\ve^2\hat T(0,0)-Z_1\tan^2\al=
\int_{T_2} \ve^2|\tau|^2 d\x-Z_1\tan^2\al=0~ 
\ee  
by the definition of $\al$ in (\re{3al}) since $\Vert\psi^0\Vert=\sqrt{Z_1}$. 
 \medskip

2) The second quotient belongs to  $L^2(-1,1)$, since 
\be\la{las} 
\fr{\hat\rho^0}{|\xi|}=\fr{{\hat\mu^{\rm per}_1}}{|\xi|}+e\fr{{\hat P}}{|\xi|}= 
\fr 
{{\hat\mu^{\rm per}_1}+eZ_1}{|\xi|}+ 
e\fr{{\hat P}-Z_1}{|\xi|}, 
\ee 
where all the functions are taken at the point $(0,\xi)$. 
Here the left-hand side belongs to $L^2(-1,1)$, 
since $\Lam\rho^0\in L^2$, while the first term on the right 
belongs to $L^2(-1,1)$ 
 by our assumption (\re{rous5}). 
\medskip\\
{\it ii)}
 The bound (\re{ltm}) holds for both terms of (\re{sc}) 
by the arguments above
since 
$\tan\al\sim\sin\al\sim\ve$ as $\ve\to 0$.\bo
\medskip

Formula (\re{3rod2}) implies (\re{Gder3}),
where the limit converges in  $L^2$ by (\re{ltm}).\bo
\bigskip\\
ii) 
Lemma \re{lL2} implies the bound (\re{HamsT223}). Formula
 (\re{Gder2})  follows by differentiation of  (\re{HamsT223}) in $\ve$.
\bo

\subsection{The variational identity} 
 
Since $\psi^0$ is a minimal point, the G\^ateaux derivative (\re{Gder2}) 
vanishes: 
\be\la{GaD} 
\int_{T_2}  \Big[\fr{\h^2}{2\cm} 
(\na\tau\ov{\na\psi^0}+\na\psi^0\ov{\na \tau}) 
+
e {\Lam\rho^0}\Lam(\tau\ov{\psi^0}+\psi^0\ov \tau )
\Big]d\x=0. 
\ee 
Substituting $i\tau$ instead of $\tau$ in this identity 
and subtracting, we obtain 
\be\la{GaD2} 
-\fr{\h^2}{2\cm}\langle\De\psi^0, \tau\rangle 
+ e \langle 
\Lam\rho^0,\Lam(\tau\ov{\psi^0} ) \rangle=0. 
\ee 
Next step 
we should evaluate the ``nonlinear'' term.

\bl For the limit functions \eqref{subs2} we have 
\be\la{sp} 
\langle 
\Lam\rho^0,\Lam(\tau\ov{\psi^0} ) \rangle= 
\langle\phi^0\psi^0, \tau \rangle, \qquad \tau\in\ccT^0,
\ee 
where $\phi^0$ is any potential satisfying \eqref{rho2}.

\el 
\Pr 
First we note that $\Lam\rho^0\in L^2$ by \re{Lamro}), and 
$\Lam(\tau\ov{\psi^0} )\in L^2$ as we have established in the proof of Lemma \re{lL2}.
Moreover, 
$\rho^0=-\De\phi^0$. 
 Then, by the Parseval--\allowbreak Plancherel identity, 
\be\la{GaD3} 
\langle 
\Lam\rho^0,\Lam(\tau\ov{\psi^0} ) \rangle= 
\sum_{\bk\in \Ga_2^*\setminus 0}\int \hat\phi^0(\bk,\xi)\ov{\widehat{\tau\ov{\psi^0} }(\bk,\xi)}d\xi+
\lim_{\ve\to 0+}
\int_{|\xi|>\ve}
\hat\phi^0(0,\xi)
\ov{\widehat{ \tau\ov{\psi^0}}(0,\xi)}d\xi= 
\langle\hat \phi^0,\widehat{\tau\ov{\psi^0} } \rangle, 
\ee 
where $\hat \phi^0$ is the distribution on $\Si_2$.
The last identity holds 
(and the right hand side is well defined)
by (\re{T0}) since $\xi\hat\phi^0(0,\xi)\in L^2(-1,1)$ due to (\re{rho2})
with $G^0\in L^2\otimes\R^3$. Finally, 
\be\la{GaD4} \langle\hat \phi^0,\widehat{\tau\ov{\psi^0} } \rangle 
= \langle\phi^0,\tau\ov{\psi^0} \rangle=\int \phi^0(\x)\ov{\tau}(\x)\psi^0(\x)d\x
\ee
by an obvious extension of the Parseval--\allowbreak Plancherel identity.\bo 
\medskip 
 
Using (\re{sp}), we can rewrite (\re{GaD2}) as the variational identity similar to (\re{3GaD22}):
\be\la{GaD22} 
\langle -\fr{\h^2}{2\cm}\De\psi^0 
+ e 
\phi^0\psi^0, \tau \rangle=0,\qquad \tau\in\ccT^0. 
\ee 
 
\subsection{The Schr\"odinger equation}

 Now we prove the Schr\"odinger equation (\re{LPS3}) with $d=2$.
\bl\la{lse2}
$\psi^0$ is the eigenfunction of the 
Schr\"odinger operator:
\be\la{Hpsi} 
H\psi^0=\lam\psi^0,
\ee 
where 
$\lam\in\R$. 
\el
\Pr
 This equation  with $\lam\in\C$ follows as in Lemma \re{lse}.
It remains to verify that $\lam$ is real. 
Our plan is standard: to multiply 
(\re{Hpsi}) by $\psi^0$ and to integrate. 
{\it Formally}, we would obtain 
\be\la{Hpsi2} 
\langle H\psi^0,\psi^0\rangle = 
\lam\langle\psi^0,\psi^0\rangle. 
\ee 
However, it is not clear that 
the left-hand side is well defined and real since the potential $\phi^0(\x)$
can grow by (\re{wfi2}).

To avoid this problem, we  multiply by a  function
$\psi_\ve\in H^1$ with compact support, where $\ve>0$ and 
$\Vert\psi_\ve-\psi^0\Vert_{H^1}\to  0$ as $\ve\to 0$. Then 
\be\la{Hpsi3} 
\langle H\psi^0,\psi_\ve\rangle = 
\lam\langle\psi^0,\psi_\ve\rangle, 
\ee 
and the right-hand side converges to  the one of  (\re{Hpsi2}) as $\ve\to 0$. 
Hence, the left-hand sides also converge.  
 In  detail, 
\be\la{Hpsi4} 
\langle H\psi^0,\psi_\ve\rangle = 
-\fr{\h^2}{2\cm}\langle \De\psi^0,\psi_\ve\rangle+ 
\langle\phi^0\psi^0,\psi_\ve\rangle. 
\ee 
For the middle term, the limit exists and is real. 
Therefore, identity (\re{Hpsi3}) 
implies that the last term is also converging, and hence
it remains to 
make its  limit  real by 
a~suitable choice of approximations $\psi_\ve$. 
We note that 
\beqn\la{cond2} 
\langle\phi^0\psi^0,\psi_\ve\rangle=\lim_{\de\to 0} 
\langle\phi^0\psi_\de,\psi_\ve\rangle 
=\lim_{\de\to 0} 
\langle\phi^0,\ov\psi_\de\psi_\ve\rangle, 
\eeqn 
since $\phi^0\in H^2_{\rm loc}(T_2)\subset C(T_2)$. 
Hence,  we can set 
\be\la{appr} 
\psi_\ve(\x)=\chi(\ve x_3)\psi^0(\x). 
\ee 
where $\chi$ is a real function from $C_0^\infty(\R^3)$ with $\psi(0)=1$. 
Now the functions $\ov \psi_\de(\x)\psi_\ve(\x)$ are real 
for all $\ve,\de>0$. It remains to note that the potential $\phi^0(\x)$ 
is also real by Lemma \re{lP2}.\bo
 \medskip
 
 This lemma implies  equation (\re{LPS3}). Therefore, $\psi^0\in H^2_{\rm loc}(T_2)$ since $\phi^0\in C(T_2)$.
 Theorem \re{tgs2} ii) is proved.

\subsection{Smoothness of ground state} 
  We have proved that 
  $\psi^0\in H^2_{\rm loc}(T_2)$ under conditions (\re{roL1}) and (\re{rous5}). 
   Using the Schr\"odinger equation (\re{LPS3}) 
  we can improve the smoothness
 of $\psi^0$ strengthening  the condition (\re{roL1}).
Namely, let us assume that
\be\la{rpem} 
\mu_1^{\rm per}\in C^\infty(T_2).
\ee 
For example,  (\re{rpem}) holds if $\mu_1\in  \cS(\R^3)$,
where $\cS(\R^3)$ is the Schwartz space of test functions.

\bl\la{lpp2} 
Let condition \eqref{rpem} hold, and 
$\psi^0\in H^2_{\rm loc}(T_2)$, $\phi^0\in H^2_{\rm loc}(T_2)$ is a solution to equations 
\eqref{LPS3}--\eqref{LPS3g} with $d=2$. Then 
the functions $\psi^0,\phi^0$ are smooth.

\el 
The proof is similar to the one of Lemma \re{lpp}.

%%%%%%%%%%%%%%%%%%%%%%%%%%%%%%%%%%%%%%%%%%%%%%%%%%%%%%%%%%%%%%%%%%%%%%%%%% 
 
\setcounter{subsection}{0} 
\setcounter{theorem}{0} 
\setcounter{equation}{0} 
 
\section{1D lattice} 
 
The case of a~one dimensional lattice 
$\Ga_1$ is very similar to the 2D case, 
though some of our  constructions and arguments require  suitable 
modifications. 
For $d=1$
we can assume  $\Ga_1=\Z$ without loss of generality and
construct a solution to system   \eqref{LPS3}--\eqref{LPS3g} 
for the corresponding functions on the `slab' 
 $T_1:=\R^3/\Ga_1=\T^1\times\R^2$ 
with coordinates $\x=(x_1,x_2,x_3)$, where 
$x_1\in \T^1$, and $(x_2,x_3\in\R^2$. 
Now we denote by $H^s$ the  complex Sobolev space on $T_1$, and 
by $L^p$, the  complex Lebesgue space of  functions on $T_1$. 
 
The existence of the ground state follows
by minimizing the energy  
(\re{HamsT}), where the integral is extended 
over $T_1$ instead of $T_3$. 
The neutrality condition 
of type (\re{neu1})
holds for 
$\Ga_1$-periodic states with finite energy, as for $d=2$. 

Again we restrict ourselves by $N=1$, 
so $\ov\x^0=\x^0_1$ can be chosen arbitrary, and we set $\x^0_1=0$.

The energy in the slab $T_1$
is defined by expression similar to (\re{HamsT22}): 
\be\la{HamsT221} 
\E(\psi):= 
\int_{T_1}\Bigl[ 
\fr{\h^2}{2\cm}|\na\psi(\x)|^2 
+ 
\fr12 |\Lam\rho(\x)|^2 
\Bigr]d\x, \qquad \rho(\x):=\si^0(\x) +e|\psi(\x)|^2. 
\ee 
Here $\si^0=\mu_i^{\rm per}\in L^1\cap L^2$ as in  (\re{rrr22}).
Hence, 
\be\la{intro21} 
\ds\int_{T_1} 
\si^0(\x)d\x=Z_1|e|,~~~~~Z_1>0. 
\ee  
Now 
the Fourier representation for the test functions 
$\vp(x)\in C_0^\infty(T_1)$
is defined by 
\be\la{FR21} 
\vp(\x)= \fr 1{2\pi} 
\sum_{\bk\in \Ga_1^*} e^{-i\bk x_1} 
\int_{\R^2} e^{-i(\xi_1x_2+\xi_2x_3)}\hat \vp(\bk,\xi)d\xi, 
\ee 
where  $\Ga_1^*=2\pi\Ga_1$
and 
\be 
\la{FT21} 
\hat 
\vp(\bk,\xi)=F\vp(\bk,\xi)=\fr 1{2\pi} 
\int_{T_1} e^{i(\bk x_1+\xi_1x_2+\xi_2x_3)} \vp(\x)d\x,\qquad  (\bk,\xi)\in  \Si_1:=\Ga_1^*\times \R^2. 
\ee 
The operator $\Lam=(-\De)^{1/2}$ is defined for $\vp\in L^1\cap L^2$ by 
the same formula (\re{Lam})
provided the quotient belongs to $L^2(\Si_1)$. This implies 
\be\la{021} 
\hat \vp(0,0)=0. 
\ee 
For $\psi\in H^1$ with finite energy (\re{HamsT221}) 
we have $\Lam\rho\in L^2(\Si_1)$, and 
hence, (\re{021}) with  $\vp=\rho$ implies 
the neutrality condition 
(\re{neu2}) with $T_1$ instead of $T_2$: 
\be\la{neu21} 
\hat\rho(0,0)=\int_{T_1} \rho(\x)d\x= 
\int_{T_1} [\si^0(\x) +e|\psi(\x)|^2]d\x=0. 
\ee 
Now (\re{intro21}) gives
\be\la{M21} 
\int|\psi(\x)|^2d\x=Z_1.
\ee 
Thus, the finiteness of the Coulomb energy 
$\Vert\Lam\rho\Vert^2$ prevents the electron charge from escaping to infinity, 
as in 2D case.

Finally, the Fourier transform $F:\psi\mapsto\hat\psi$ 
is a unitary operator from $L^2(T_1)$ to $L^2(\Si_1)$. Hence, 
energy (\re{HamsT22}) reads 
\be\la{HamsT22F1} 
\E(\psi)= 
\sum 
_{\bk\in \Ga_1^*} 
\int_{\R^2}\Bigl[ 
\fr{\h^2}{2\cm}(\bk^2+\xi^2)|\hat\psi(\bk,\xi)|^2 
+ 
\fr12 
\fr {|\hat\rho(\bk,\xi)|^2} {\bk^2+\xi^2} 
\Bigr]d\xi. 
\ee 
 
\bd 
$M_1$ denotes the set of  
$\psi\in H^1$ satisfying the neutrality condition (\re{M21}). 
\ed 
 
We note that 
(\re{intro21}) can be written as
$\hat\mu^{\rm per}_1(0)+eZ_1=0$. 
We assume  moreover, 
\be\la{rous51} 
\mbox{\bf Condition III.}\qquad
\fr{\hat\mu^{\rm per}_1(0,\xi)+eZ_1}{|\xi|}\in L^2(D),\qquad
D:=\{\xi\in\R^2: |\xi|\le 1\} 
\qquad
\ee 
similarly to (\re{rous5}).
For example, this condition holds, 
provided that  
\be\la{rous5if1} 
\int_{\R^3}(1+|x_2|+|x_3|)|\mu_1(\x)|d\x <\infty.
\ee 
The third main result of the present paper is the following.
 
 \bt\la{tgs1} 
Let conditions \eqref{roL1} and \eqref{rous51} hold,  and $N=1$. Then 
\medskip\\
i) There exists 
$\psi^0\in M_1$ with
\be\la{U0min1} 
\E(\psi^0)=\inf_{\psi\in M_1}\E(\psi).
\ee 
ii) Moreover, $\psi^0\in H^2_{\rm loc}(T_1)$ and
satisfies equations \eqref{LPS3}--\eqref{LPS3g} with $d=1$,
where the
potential $\phi^0\in H^2_{\rm loc}(T_1)$ is real, $\x^0_1=0$, 
 and $\om^0\in\R$. 
\medskip\\
iii) The following bound holds
\be\la{wfi1} 
|\phi^0(\x)|\le 
C(1+|x_2|+|x_3|)^{1/2},\qquad\qquad \x\in T_2.
\ee 
\et

The proof is similar to the one of Theorem \re{tgs2}.
As in 2D case, we obtain 
$\psi^0\in M_1$ as a minimizer for the energy (\re{HamsT221}).
The  potential 
$\phi^0$ can be constructed by a modification of  Lemma \re{lP2}, see Appendix below.
 \medskip

 Finally, next lemma follows similarly to Lemma \re{lpp}.

 \bl\la{lpp1} 
 The functions $\psi^0,\phi^0$ are smooth 
under condition 
\be\la{rpem1} 
\mu_1^{\rm per}\in C^\infty(T_1).
\ee 

\el

 \appendix
 
 \setcounter{section}{0}
\setcounter{equation}{0}
\protect\renewcommand{\thesection}{\Alph{section}}
\protect\renewcommand{\theequation}{\thesection. \arabic{equation}}
\protect\renewcommand{\thesubsection}{\thesection. \arabic{subsection}}
\protect\renewcommand{\thetheorem}{\Alph{section}.\arabic{theorem}}

 \section{The potential of 1D lattice} 
 We start with obvious modifications of the proof of Lemma \re{lP2}.
 Namely, 
 the potential $\phi^0(\x)$ for the 1D lattice 
 satisfies the equation of type (\re{rho2}) with 
 \be\la{G0}
 G^0:= 
-  iF^{-1}\fr{(\bk,\xi)}{\bk^2+\xi^2}\hat\rho^0(\bk,\xi)
 \in L^2(T_1), \qquad \rot G^0(\x)\equiv 0.
 \ee
 We use  the splitting of type  (\re{split}), and
 respectively, the solution splits as $\phi^0=\phi_1+\phi_2$. 
The second solution 
$\phi_2\in H^2$  as in the proof of Lemma \re{lP2}.
Hence, $\phi_2$
is bounded continuous function on $T_1$ by the 
Sobolev embedding theorem.
 
 On the other hand, the analysis of the first solution needs some 
 modifications.
Now $G_1(\x)=g_1(x_2,x_3)\in L^2(\R^2)\otimes \R^2$
 is the real vector field,
and $\supp \hat g_1\subset \{\xi\in\R^2: |\xi|\le 1\}$.
Therefore,
$g_1$ is the smooth function, and 
\be\la{Dphi1}
 \De \phi_1=\na\cdot g_1\in L^2(\R^2), \qquad \rot g_1(\x)\equiv 0.
 \ee
  Respectively, the solution to 
$\na \phi_1= g_1$
is given by 
the contour integral
\be\la{pott} 
\phi_1(\x)=\int_0^\x g_1(\y)d\y+C,\qquad\qquad \x\in \R^2,
\ee 
which does not depend on the path in $\R^2$. This solution is real and smooth.
\medskip

We still should prove the estimate (\re{wfi1}).
We will 
deduce it from the corresponding 
estimate 'in the mean'. 
Let us denote the circle $B:=\{\x\in\R^2:|\x|<1\}$.

\bl\la{lpot}
For any unit vector 
$\e\in\R^2$ 
\be\la{esm}
\Vert\phi_1 \Vert_{L^2(B+\e R)}\le C(1+R)^{1/2},\qquad \qquad R>0.
\ee
\el
\Pr  First, (\re{pott}) implies that 
\be\la{pot2} 
\phi_1(\x+\e R)-\phi_1(\x)=\int_0^R g_1(\x+t\e)dt, \qquad\qquad \x\in \R^2
\ee 
for any $R\in\R$. 
Now
the Cauchy-Schwartz inequality implies that
\be\la{pot23} 
|\phi_1(\x+\e R)|^2\le C_1+2R\int_0^R |g_1(\x+t\e)|^2dt, \qquad\qquad \x\in B
\ee 
since  the function $\phi_1$ is bounded in $B$.
Finally, averaging over $\x\in B$, we get 
\be\la{pot24} 
\int_B|\phi_1(\x+\e R)|^2d\x\le C_1|B|+2R\int_0^R \int_B|g_1(\x+t\e)|^2d\x \5 dt\le C_1+C_2R\Vert g_1\Vert_{L^2(\R^2)}^2.
\ee 
Hence, 
 (\re{esm})
is proved.\bo
\medskip

Now  (\re{wfi1}) follows from the Sobolev embedding theorem:
\be\la{sapr}
\max_{\x\in B+\e R}|\phi_1(\x)|\le C_3   \Vert\phi_1 \Vert_{H^2(B+\e R)}\le 
C_4[ \Vert\De\phi_1 \Vert_{L^2(B+\e R)}+ \Vert\phi_1 \Vert_{L^2(B+\e R)})]\le C(1+R)^{1/2}
\ee
since $\De \phi_1\in L^2(\R^2)$ by (\re{Dphi1}). 
\medskip

\br
Our estimate  (\re{wfi1}) seems to be  far from optimal since
 the potential of uniformly charged line grows logarithmically  with the distance,
 % and the potential of the one-dimensional lattice with alternating charges decays like 
 % 1/distance.
One could expect an optimal estimate 
$$
|\phi^0(\x)|\le C[\log(2+|x_2|+|x_3|)]^{1/2}
$$
in the case $\na\phi^0\in L^2$
due to the example
$
\phi(\x)=[\log(2+|x_2|+|x_3|)]^{1/2-\ve}
$
with $\na \phi(\x)\in L^2$ for  $\ve>0$.

\er


\begin{thebibliography}{99} 
 
 \bibitem{Adams}
  R.A. Adams, J.J. Fournier, J. F. John, Sobolev Spaces,
  Elsevier/Academic Press, Amsterdam, 2003.
 
 
 
\bibitem{Ben02} K.~Benmlih, 
Stationary solutions for a Schr\"odinger--\allowbreak Poisson  system in 
$\R^3$, {\em Electron. J. Differ. Equ.} {\bf Conf. 09} 
(2002), 65--76. 
 
http://ejde.math.swt.edu or http://ejde.math.unt.edu 
 
 
\bibitem{BBL2003} 
X. Blanc, C.~Le Bris, P.\,L.~Lions, 
A definition of the ground state energy for systems 
composed of infinitely many particles, 
{\em Comm. Partial Diff. Eq.} {\bf 28} (2003), no. 1/2, 439--475. 
 
\bibitem{BBL2007} 
X. Blanc, C. Le Bris, P.\,L.~Lions, 
The energy of some microscopic stochastic lattices, 
{\em Arch. Ration. Mech. Anal.} {\bf 184} (2007), no.~2, 303--339. 
 
 
\bibitem{BL2005}	 
C. Le Bris, P.\,L.~Lions, 
From atoms to crystals: a mathematical journey, 
{\em Bull. Am. Math. Soc., New Ser.} 
{\bf  42} (2005), no. 3, 291--363. 
 
 
\bibitem{BO} 
M. Born, J.\,R.~Oppenheimer, Zur Quantentheorie der Molekeln, 
{\em Annalen der Physik} {\bf 389 (20)} (1927), 457--484. 
% doi:10.1002/andp.19273892002 
 
\bibitem{CB} 
E. Canc\`es,  C.~Le Bris, 
On the time-dependent Hartree--\allowbreak Fock equations coupled with a classical nuclear dynamics,  {\em Math. Models Methods Appl. Sci.} {\bf 9} (1999), no.7, 963--990. 
 
\bibitem{CL2010} 
E. Canc\`es, M. Lewin, 
The electric permittivity of crystals in the reduced Hartree--\allowbreak Fock approximation, 
{\em Arch. Ration. Mech. Anal.} {\bf 197} (2010), 139--177. 
 
%%%%%%%%% HF 
\bibitem{CS2012} 
E. Canc\`es, G. Stoltz, 
A mathematical formulation of the random phase approximation for crystals, 
{\em Ann. Inst. Henri Poincar\'e, Anal. Non Lin\'eaire} {\bf 29} (2012), 887--925. 
 
\bibitem{CBL1996} 
L. Catto, C. Le Bris, P.\,L.~Lions, 
Thermodynamic limit for Thomas-Fermi type models, 
{\em C. R. Acad. Sci., Paris, S\'er.} I 
{\bf 322} (1996), no.4, 357--364. 
 
\bibitem{CBL1998} 
L. Catto, C. Le Bris, P.\,L.~Lions, 
  The Mathematical Theory of Thermodynamic Limits: Thomas--\allowbreak Fermi Type Models, 
Clarendon Press, Oxford, 1998. 
 
\bibitem{CBL2001} 
L. Catto, C. Le Bris, P.\,L.~Lions, 
 On the thermodynamic limit for Hartree--\allowbreak Fock type models, 
{\em Ann. Inst. Henri Poincar\'e, Anal. Non Lin\'eaire} 
{\bf 18} (2001), no. 6, 687--760. 
 
\bibitem{CBL2002} 
L. Catto, C. Le Bris, P.\,L.~Lions, On some periodic Hartree-type models for crystals, 
{\em Ann. Inst. Henri Poincar\'e, Anal. Non Lin\'eaire} 
{\bf 19} (2002), no.2, 143--190. 
 
\bibitem{Dyson1967} 
F.J. Dyson, A. Lenard, Stability of matter, I, 
{\em J. Math. Phys.} {\bf 8} (1967), 423--434. 
 

%\bibitem{Fowler} M. Fowler,
%Electrons in one dimension: the Peierls transition.
%http://galileo.phys.virginia.edu/classes/752.mf1i.spring03/PeierlsTrans.htm



\bibitem{GLL2007} 
A. Giuliani, J.\,L.~Lebowitz, E.\,H.~Lieb, 
Periodic minimizers in 1D local mean field theory, 
{\em Comm. Math. Phys.} {\bf 286} (2009), 163--177. 
 

\bibitem{Kawohl2012} 
B. Kawohl, S. Kr\"omer, 
Uniqueness and symmetry of minimizers of Hartree 
type equations with external Coulomb potential, 
{\em Adv. Calc. Var.} {\bf 5} (2012), no.~4, 427--432. 
 
\bibitem{Lemm} 
 M.C. Lemm, Stability of Matter, Ph.D., LMU,  2010.
 
http://www.mathematik.uni-muenchen.de/~lerdos/Stud/lemm.pdf 
 
%\bibitem{LJ1924} 
%J.\,E.~Lennard-Jones, 
%On the determination of molecular fields. II. 
%From the Equation of State of a Gas, 
%{\em Proc. R. Soc. Lond. A} {\bf 106} (1924), 463--477. 
% doi: 10.1098/rspa.1924.0082 
 
 \bibitem{LS2014-1} 
M. Lewin, J. Sabin,
The Hartree equation for infinitely many particles. I. Well-posedness theory,  arXiv:1310.0603.

\bibitem{LS2014-2} 
M. Lewin, J. Sabin,
The Hartree equation for infinitely many particles. II. Dispersion and scattering in 2D,  arXiv:1310.0604.

 
 
 
 
\bibitem{Lieb2005} 
   E.H. Lieb, 
  The Stability of Matter: From Atoms to Stars. 
Selecta of Elliott H. Lieb. 4th ed., 
Springer,  Berlin, 2005. 
 
 
\bibitem{LL1972} 
E.H. Lieb, J.L. Lebowitz, 
The constitution of matter: 
existence of thermodynamics for systems composed of 
electrons and nuclei, {\em Adv. Math.} {\bf 9} (1972), no.~3, 316--398. 
 
\bibitem{LL2005} 
E.H. Lieb, M. Loss, 
The thermodynamic limit for matter interacting with 
Coulomb forces and with the quantized electromagnetic field: 
I. The lower bound, 
{\em Comm. Math. Phys.} {\bf 258} (2005), 675--695. 
 
\bibitem{Lieb2009} 
E.H. Lieb, R. Seiringer, The Stability of Matter in Quantum Mechanics, 
Cambridge University Press, Cambridge, 2009. 
 
\bibitem{LS1977} 
E.H. Lieb, B. Simon, 
The Hartree--\allowbreak Fock theory for Coulomb systems, 
{\em Comm. Math. Phys.} {\bf 53} (1977), 185-194. 
 
\bibitem{LAK}
 I.M. Lifshits, M.Ya. Azbel, M.I. Kaganov,
Electron Theory of Metals,
Consultants Bureau,
New York, 1973. 
 
 
 
\bibitem{Lions1981} 
P.L. Lions,  Some remarks on Hartree equation, 
{\em Nonlinear Anal., Theory Methods Appl.} 
{\bf 5} (1981), 1245--1256. 
 
\bibitem{Lions1987} 
P.-L. Lions, 
Solutions of Hartree--\allowbreak Fock equations for Coulomb systems, 
{\it Commun. Math. Phys.} {\bf  109} (1987), 33--97. 
 
\bibitem{Nier93} 
F. Nier, 
Schr\"odinger--\allowbreak Poisson  systems in dimension $d\le 3$: 
The whole-space case, 
{\em Proc. R. Soc. Edinb., Sect. A} 
{\bf 123} (1993), no.~6, 1179--1201. 
 


\bibitem{Peierls} 
R.E. Peierls,
Quantum Theory of Solids, Clarendon Press, Oxford, 2001.



\bibitem{Sob} 
S.L. Sobolev, 
Some Applications of Functional Analysis in Mathematical Physics,
AMS, Providence, RI, 1991. 
 
\end{thebibliography}
\end{document}